\documentclass[%
 aip,
 amsmath,amssymb,
 reprint,%
]{revtex4-1}

\usepackage{graphicx}
\usepackage{dcolumn}
\usepackage{bm}

\usepackage[utf8]{inputenc}
\usepackage[T1]{fontenc}
\usepackage{mathptmx}

\usepackage{balance}
\usepackage{times,mathptmx}
\usepackage{graphicx} 
\usepackage{lastpage}
\usepackage{float}
\usepackage{fancyhdr}
\usepackage{fnpos}
\usepackage[english]{babel}
\addto{\captionsenglish}{%
  
}
\usepackage{array}
\usepackage{droidsans}
\usepackage{charter}
\usepackage[usenames,dvipsnames]{xcolor}

\begin{document}

\preprint{AIP/123-QED}

\title{Characterization of MIPS in a suspension of repulsive Active Brownian Particles through dynamical features}

\author{Jos\'e Martin-Roca}
\affiliation{Departamento de Estructura de la Materia, F\'isica T\'ermica y Electr\'onica, Universidad Complutense de Madrid, 28040 Madrid, Spain}
\affiliation{GISC - Grupo Interdisciplinar de Sistemas Complejos 28040 Madrid, Spain}
\author{Raul Martinez }
\affiliation{Departamento de Estructura de la Materia, F\'isica T\'ermica y Electr\'onica, Universidad Complutense de Madrid, 28040 Madrid, Spain}
\affiliation{GISC - Grupo Interdisciplinar de Sistemas Complejos 28040 Madrid, Spain}
\affiliation{Departamento de F\'isica Te\'orica de la Materia Condensada, Facultad de Ciencias, Universidad Aut\'onoma de Madrid, 28049 Madrid, Spain}
\author{Lachlan C. Alexander}
\affiliation{Physical and Theoretical Chemistry Department, University of Oxford, United Kingdom.}
\author{Angel Luis Diez }
\affiliation{Departamento de Estructura de la Materia, F\'isica T\'ermica y Electr\'onica, Universidad Complutense de Madrid, 28040 Madrid, Spain}
\affiliation{GISC - Grupo Interdisciplinar de Sistemas Complejos 28040 Madrid, Spain}
\author{Dirk G. A. L. Aarts}
\affiliation{Physical and Theoretical Chemistry Department, University of Oxford, United Kingdom.}
\author{Francisco Alarcon}
\affiliation{ Departamento de Estructura de la Materia, F\'isica T\'ermica y Electr\'onica, Universidad Complutense de Madrid, 28040 Madrid, Spain}
\affiliation{Departamento de Ingenier\'ia F\'isica, Divisi\'on de Ciencias e Ingenier\'ias, Universidad de Guanajuato, Loma del Bosque 103, 37150 Le\'on, Mexico.}
\author{Jorge Ram{\'i}rez}
\affiliation{Departamento de Ingenier{\'i}a Qu{\'i}mica, ETSI Industriales, Universidad Polit{\'e}cnica de Madrid, 28006 Madrid, Spain}
\author{Chantal Valeriani}
\affiliation{Departamento de Estructura de la Materia, F\'isica T\'ermica y Electr\'onica, Universidad Complutense de Madrid, 28040 Madrid, Spain}
\affiliation{GISC - Grupo Interdisciplinar de Sistemas Complejos 28040 Madrid, Spain}

\date{\today}

\begin{abstract}
{\color{black}We study a two-dimensional system composed by Active Brownian Particles (ABP), focusing on the onset of Motility Induced Phase Separation(MIPS), by means of molecular dynamics simulations. For a pure hard-disk system with no translational diffusion, the phase diagram would be completely determined by their density and Péclet number. In our model, two additional effects are present: traslational niose and the overlap of particles; we study the effects of both in the phase space. As we show, the second effect can be mitigated if we use, instead of the standard Weeks-Chandler-Andersen potential, a stiffer potential, the pseudo-hard spheres potential. Moreover, in determining the boundary of our phase space, we explore different approaches to detect MIPS and conclude that observing dynamical features, via the non-Gaussian parameter, is more efficient than observing structural ones, such as through the local density distribution function. We also demonstrate that the Vogel-Fulcher equation successfully reproduces the decay of the diffusion as a function of density, with the exception of very high densities. Thus, in this regard, the ABP system behaves similarly to a fragile glass.}
\end{abstract}

\maketitle

\section{\label{sec:level1}Introduction}


Active matter is a branch of physics that studies out-of-equilibrium systems in which energy is supplied at the level of individual entities called active particles.  Active particles dissipate energy while performing motion \cite{ramaswamy2017active, bechinger2016active}. Such out of equilibrium behaviour can result in interesting collective phenomena not observed in equilibrium systems. Examples of collective motion of active living systems can be found at every length scale: from colonies of bacteria, to flocks of birds  \cite{Vicsek_Rev}. More recently, similar collective behavior has been experimentally mimicked by synthetic active colloids \cite{reviewdauchot,Aronson,volpe2014,narinder2019active, narinder2018memory,berg1993random, ginot2018sedimentation,patteson2015running, aragones2018diffusion}, whose activity can be tuned either magnetically \cite{xu2020,calero}, chemically \cite{palacci2010,ginot2015,campbell2019}, via electric fields (thanks to induced-charge electrophoresis effects)\cite{sprenger2020,Granick,dauchot2019} or by means of UV light illumination \cite{xu2020,rao2020,palacci2014}.

 On the theoretical side, models have been developed to get a better understanding of the physics of active matter, that might allow for the tailoring new smart materials. One of the simplest yet most insightful models is that of the so-called Active Brownian Particles (ABPs), where Brownian dynamics equations of motion have been modified to allow for particle self-propulsion and a gradual change in direction (tuned by rotational diffusion)\cite{cates2013,pagonabarraga2018,cugliandolo,gonnella1}. One paradigmatic phenomenon observed in suspensions of repulsive ABPs is the emergence of Motility Induced Phase Separation (MIPS). Despite the lack of explicit attractive forces between particles, particles tend to phase separate into a dense region where they move slower and a dilute region where they move faster. MIPS, which can appear in either two \cite{FilyPRL,RednerPRL,Stenhammar,Bialke,SpeckBinder} or three dimensions \cite{Stenhammar,Wysocki}, has been detected by computing the local density (a structural feature). Interesting questions have emerged from these studies, one of which is the role played by the active force on the particles' effective diameter, as briefly discussed in a previous work\cite{Stenhammar} where the Weeks-Chandler-Andersen (WCA) potential was used. The relative strength of the self-propulsion force and the repulsive inter-particle interaction force determines the distance between the particles in the denser regions of the simulation, as WCA allows some degree of overlap. Some works, including studies of ABPs, have avoided overlap by using harder interaction potentials\cite{Voigtmann_JPCM18,PRL_Levis,LevisCodina,digregorio20192d}.
The effect of the nature of the repulsive interaction on systems in thermodynamic equilibrium has been thoroughly studied \cite{auer2002crystallization,hynninen2003phase,prestipino2005phase,taffs2013structure,PhysRevA.4.1597,Phase_Behavior_of_Concentrated,bialke2012crystallization,filion2010crystal, pieprzyk2019thermodynamic, filion2011simulation}. In this work, we provide a closer look on the effects determining the shape of the MIPS phase space and the importance of the softness of the potential. 

The goal of our work is two-fold. (i) To study the influence of different parameters (besides P\'eclet number and density) in the onset of MIPS, and compare two potentials of different softness: the often used but relatively soft WCA\cite{WCA} and a stiffer Pseudo-hard sphere potential (PHS)\cite{PHS}, a continuous potential that mimics structural and dynamical properties of  hard-spheres in equilibrium\cite{PHS,espinosa}, and out-of-equilibrium\cite{rosales,montero}. (ii) To study the system's dynamics, we propose the use of dynamical properties, instead of structural ones, to locate the appearance of MIPS in the state diagram. In addition, we analyse the suitability of the Vogel-Fulcher equation for the description of the dynamical behavior of an ABP suspension, as compared to its passive analogue.

The manuscript is organized as follows: In section 2 we present the simulation details, in section 3 we show and discuss the results and in section 4 we present our conclusions.

\section{\label{sec:level1}Simulation details}

The simulated system consists of $N=20000$ circular particles with diameter $\sigma$ in a two-dimensional box of size $L_x \times L_y$  (where periodic boundary conditions have been implemented).  
$L_x$ and $L_y$ have been set in order to obtain the desired total density $\rho=  \frac{N}{L_x \, L_y} $, with a ratio \textcolor{black}{$L_y/L_x \approx 0.58$}. As in Ref.\cite{rogel}, we use the total density of the system instead of the packing fraction, 
 {\color{black}{since a particle's diameter (needed to compute packing fraction) might not be uniquely defined due to particles activities (and cannot be estimated via the Barker and Henderson's approach\cite{barker}).}}
 As an initial configuration, we prepare the system in a hexagonal lattice. All simulations have been run at a given density until a steady state is reached.

To simulate Active Brownian Particles, we perform Brownian Dynamics simulations with an in house modified version of the {\it LAMMPS}\cite{LAMMPS} open source package.
The equations of motion for the position $\vec{r}_i$ and orientation $\theta_i$ of the $i$-th active particle can be written as:
\begin{align}
\label{eq:motion}
& \dot{\vec{r}}_i = \frac{D_t}{k_B T} \left( - \sum_{j\neq i} \nabla V(r_{ij}) +  F_a \, \vec{n}_i \right) + \sqrt{2D_t} \, \vec{\xi}_i, \\
& \dot{\theta}_i = \sqrt{2D_r}\, \xi_{i,\theta},
\end{align}
where $V(r_{ij})$ is the inter-particle pair potential, $k_B$ the Boltzmann constant, $T$ the absolute temperature, $F_a$ a constant self-propulsion force acting along the 
orientation vector $\vec{n}_i$, which forms an angle $\theta_i$ with the positive $x$-axis, $D_t$ is the translational and $D_r$ the rotational diffusion coefficient. Furthermore, the components of the thermal forces $\vec{\xi}_i$ and $\xi_{i,\theta}$ are white noise with zero mean and correlations $\langle\xi^{\alpha}_i(t)\xi^{\beta}_j(t')\rangle = \delta_{ij} \delta_{\alpha\beta} \delta(t-t')$, where $\alpha,\beta$ are the $x$, $y$ components, and $\langle\xi_{i,\theta}(t)\xi_{j,\theta}(t')\rangle = \delta_{ij} \delta(t-t')$. 
In equilibrium, the translational, $D_t$, and rotational diffusion coefficient, $D_r$,
follow a Stokes-Einstein relation for spherical particles (with diameter $\sigma$)
\cite{StokesEinstein}: $D_r = 3D_t/\sigma^2$.
 \textcolor{black}{However, when explicitly stated, we uncouple both diffusion coefficients. In an active matter system, this uncoupling is experimentally justified\cite{berg1993random, ginot2018sedimentation,patteson2015running, aragones2018diffusion} and has also been used in a previous work\cite{fily2012athermal}}. 
 \textcolor{black}{Although we have not found in the scientific literature a specific experimental system where $D_r$ and $D_t$ can be independently modified, we argue that there is no theoretical reason to affirm that this cannot be done in an active system, so this decoupling can also be of experimental interest.}


 \begin{figure}[h!]
\centering
    \includegraphics[width=0.8 \columnwidth]{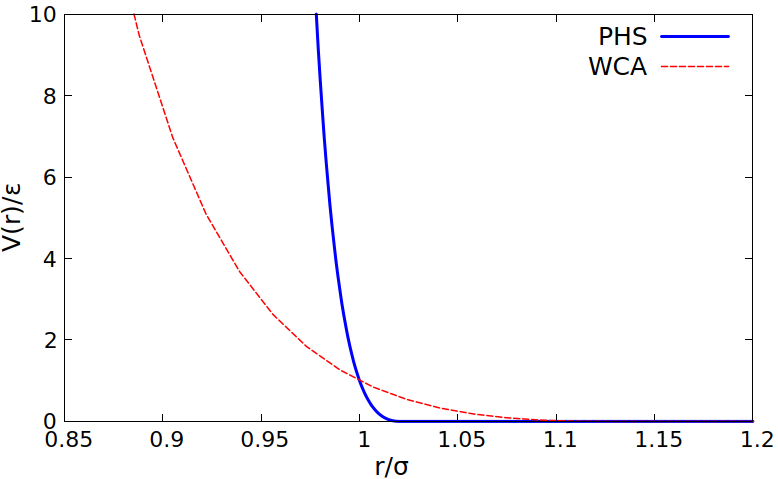} 
\caption{Repulsive potentials considered in this work. Note that WCA is softer than PHS. \label{fig:Potentials}}
\end{figure}

Throughout our study, we will consider two repulsive interaction potentials (Fig.\ref{fig:Potentials}), WCA \cite{WCA} potential (in red):
\begin{equation}
    V_{WCA}(r)  = \left\{ \begin{array}{lcl}
        4 \epsilon \left[ \left( \dfrac{\sigma}{r} \right)^{12} -  \left(\dfrac{\sigma}{r} \right)^{6} \right] + \epsilon & , & r< 2^{1/6}  \sigma \\
        0 & , & r \geq 2^{1/6}  \sigma
    \end{array}  \right.
     \label{eq:WCA}
\end{equation}
where $r$ is the center-to-center distance and $\sigma$ is the particle diameter; and the so-called pseudo-hard sphere (PHS) \cite{PHS} potential (in blue):
\begin{equation}
    V_{PHS}(r)  = \left\{ \begin{array}{lcl}
        50 \left( \frac{50}{49} \right)^{49} \epsilon \left[ \left( \dfrac{\sigma}{r} \right)^{50} -  \left(\dfrac{\sigma}{r} \right)^{49} \right] + \epsilon & , & r< \left(\frac{50}{49}  \right)  \sigma \\
        0 & , & r \geq \left(\frac{50}{49}  \right)  \sigma
    \end{array}  \right.
     \label{eq:PHS}
\end{equation}
which has been shown to properly reproduce equilibrium\cite{espinosa} and out-of-equilibrium\cite{rosales,montero}  features of a hard spheres suspension.

\textcolor{black}{Throughout this paper, all quantities are expressed in reduced units, in which lengths, times and energies are given in terms of $\sigma$, {$\tau_{LJ} = \sqrt{m\sigma /\varepsilon}$} and $\epsilon$ respectively. In all our simulations we set $\epsilon=1$, $D_t \tau_{LJ} /\sigma^2 =1.5$ (value that will be clarified below) and  the relation $D_t \, \tau_{LJ}/\sigma^2 = k_B \, T/\epsilon$.
The time step is set to $\Delta t = 10^{-5} \tau_{LJ}$ for WCA, and $\Delta t = 10^{-6} \tau_{LJ}$ for PHS \cite{Stenhammar}. We run the simulations for $10^7$ steps for WCA and $10^8$ steps for PHS to equilibriate the system. Then, we simulate for another $10^8$ steps for WCA and $10^9$ steps for PHS. All our simulations are run for $t/\tau_{LJ}=10^3$.}  \\


As a measure of the degree of activity, we use the P\'eclet number $\mathrm{Pe}$, i.e. the dimensionless ratio between advective and diffusive transport, defined as: 
\begin{equation}
\mathrm{Pe} = \frac{3 v \, \tau_r}{\sigma}= \frac{3 F_a D_t}{\sigma k_B T D_r},
\label{eq:Pe}
\end{equation}
where $v=F_a D_t/k_BT$ is the self-propelling velocity and $\tau_r = 1/D_r$, the reorientation time.

Firstly, in order to detect phase separation we follow Ref.\cite{Stenhammar} and 
compute the local density $\rho_0$ for each particle as the inverse of the area of the polygon associated to it via a Voronoi tessellation. The overall density $\rho$ is the mean of $\rho_0$ averaged over all particles. To establish if the system phase separates into a dense and a dilute phase (MIPS), we calculate the probability distribution function of the local density $P(\rho_0)$ (once the system is in the stationary state). Typically, when the system is homogeneous, $P(\rho_0)$ is characterised by a single maximum around the system's average density. When MIPS  occurs, $P(\rho_0)$ exhibits two maxima, one centered at the value of the local density characterizing the dilute phase and the other one corresponding to the concentrated phase. While the two $P(\rho_0)$ peaks are well defined deep into the MIPS region, they are not easily distinguishable near the boundary of the MIPS region.

For a better understanding of the local structure of dense phases, we compute the hexatic order parameter $\psi_6$ for the $k$-th particle in two dimensions, as:
\begin{equation}
\psi_6(k) = \frac{1}{n} \, \sum_{j\in N_k} \, \text{e}^{i \, 6 \, \theta_{kj}},
\label{eq:Psi6}
\end{equation}
where $\theta_{kj}$ is the angle between the vector $\vec{r}_{kj}$ and the x-axis, $N_k$ is the set of first Voronoi neighbors\cite{ginelli} of particle $k$ and $n$ is the number of neighbors. 
\textcolor{black}{To compute the global crystalline order, we sum the value of $\psi_6(k)$ over all particles in the system.}

As an alternative method to identify MIPS, besides computing the local density $P(\rho_0)$, we propose a method based on dynamical, rather than structural, properties. \textcolor{black}{When MIPS starts, the system is characterised by dense regions of slow particles and dilute regions of fast particles, resembling dynamic heterogeneities observed in supercooled liquids\cite{berthier-biroli}. Therefore, } we investigate the dynamical heterogeneity appearing in the system by means of the non-Gaussian parameter { which in a two dimensional system is:} 
\begin{equation}
\alpha_2(t) = \frac{\left\langle \Delta r^4(t) \right\rangle}{2 \, \left\langle \Delta r^2(t) \right\rangle^2}-1,
\label{eq:nongauss}
\end{equation}
where $\left\langle \Delta r^k(t) \right\rangle$ is the $k$-th moment of the probability distribution function (PDF) of particle displacements in two dimensions. This parameter is a measure of the deviation between the PDF and a Gaussian distribution, characteristic of pure Brownian motion. \textcolor{black}{The non-Gaussian parameter has been previously used to describe anomalous and/or heterogeneous transport dynamics in systems at equilibrium or approaching the glassy state \cite{kumar2006nature, song2019transport, vorselaars2007non, weeks2000three, ramirez2018molecular}}.
In this work, for each simulation, we define $\sigma_\alpha$ as the characteristic size of the fluctuations of $\alpha_2$ at equilibrium for the passive case at the same density. Thus, the proposed criterion to identify MIPS is the following: 
1) we compute the time average of $\alpha_2$; 2) when this average is smaller than $10 \cdot \sigma_\alpha$, 
the system is in a homogeneous phase; 
3) when this average is larger than $10 \cdot \sigma_\alpha$,the system is in a MIPS state. 
 The suitability of this method is discussed in section III.C. \textcolor{black}{For further details on the meaning of the parameter $\sigma_{\alpha}$ and how to estimate its value for each system, please see the Supplementary Info.} 

\textcolor{black}{
To better unravel the system's dynamical features, 
we also compute the long time effective diffusion coefficient $D_\mathrm{eff}$ from the slope of the mean-square displacement at long times. 
Inspired by the dynamical features of a metastable fluid, }
we try to fit the $D_\mathrm{eff}$ with the Vogel-Fulcher model (as in Ref.\cite{Ludovic})
\begin{equation}
D_\mathrm{eff}(\rho)= \exp\left(A+ \frac{B}{\rho -\rho'}\right),
\end{equation}
an empirical law frequently used in the study of glassy dynamics
\cite{morley}. 

\section{Results}


\subsection{\bf Density versus activity state diagrams}

The phase space of an ABP suspension is typically  characterised by two variables: the total density $\rho$ and the P\'eclet number, $\mathrm{Pe}$.
\begin{figure}[h!]
\centering
\includegraphics[width=1.0\columnwidth]{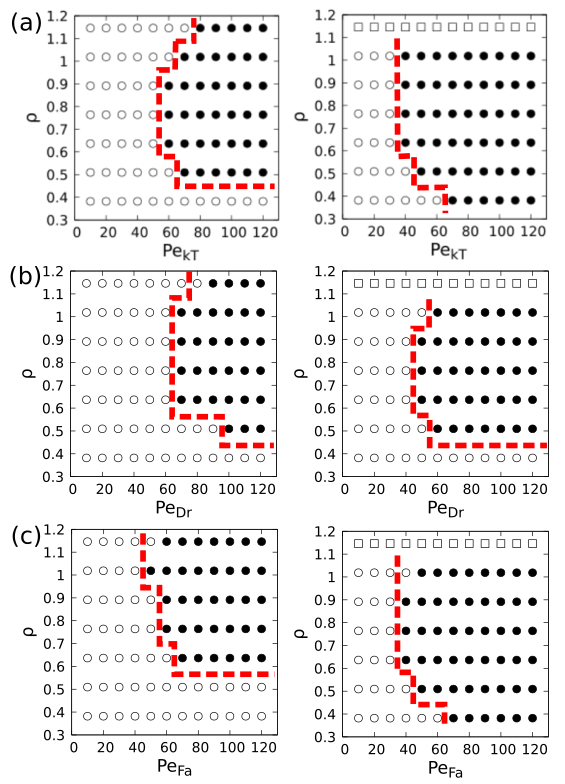}
\caption{$\rho$ versus $\mathrm{Pe}$ state diagrams of ABP interacting by means of WCA (left) and PHS (right) potentials. $\mathrm{Pe}$ is modified by varying (a) $k_B T$ while keeping $D_r=4.5$, $D_t=1.5$ and $F_a\sigma/\epsilon=24$,(b) by varying $D_r$ while keeping $k_BT/\epsilon = 1.5$, $D_t=1.5$ and $F_a\sigma/\epsilon=24$ fixed 
 and (c) by varying $F_a$ while keeping $k_BT/\epsilon=1.5$ with $D_t=1.5$ and $D_r$ coupled. Filled symbols correspond to MIPS and empty symbols to the homogeneous phase 
 determined by $P(\rho_0)$ \cite{Stenhammar}. Red dashed lines represent MIPS boundaries determined by means of the non-Gaussian parameter $\alpha_2$ (see main text).
\label{fig:WCADr}}
\end{figure}
 \textcolor{black}{
However, according to equation (\ref{eq:Pe}),  $\mathrm{Pe}$ can be varied by changing the temperature $T$ (as in Ref.\cite{Stenhammar}), the self-propelling force $F_a$ (closely resembling  experiments\cite{ginot2018}) or the rotational diffusion coefficient $D_r$ (as in Ref.\cite{prymidis}).  In this work, we attempt to quantify how the system phase behaviour changes when different approaches are used to modify $\mathrm{Pe}$.}\\

 \textcolor{black}{
In figure \ref{fig:WCADr}, we present three state diagrams for both WCA (left-hand side) and PHS (right-hand side) potentials, obtained: (a) changing $k_BT$ (while keeping $\epsilon=1$ and all other parameters fixed), (b) changing $D_r$  (while leaving all other parameters fixed, including  $D_t$, thus not coupling the diffusion coefficients via the Stokes-Einstein relationship), and (c) changing $F_a$ (while keeping all other parameters fixed).}
\textcolor{black}{ A figure where all diagrams are overlaid can be found in the Supplementary Info.}
 \textcolor{black}{
In each panel, the system undergoes motility induced phase separation (MIPS, filled symbols) when both density and activity (P\'eclet number) increase, and it can be found in an homogeneous phase at low density/activity (empty symbols). The same results have been obtained characterising the different states via the local density (symbols) or non-Gaussian parameter $\alpha_2$ (red dashed-lines, discussed in Sec.B). 
By visual inspection, we can already state that the different parameters chosen to vary the P\'eclet mostly affect the boundaries of the MIPS phase, rather than its bulk.} 
\footnote{ At very high densities PHS particles barely move. Therefore, one should make sure to prepare an initial configuration where almost no crystalline particles are present. At such high densities, one should resort to special numerical techniques (such as the one by Ref.\cite{Lubachevsky}) to prepare an initial configuration in an amorphous phase. Eventually, the amorphous solid might crystallise. As with PHS, a crystalline phase might also be detected for WCA at densities even closer to the closed packed state (since particles are allowed to partially overlap). However, studying the crystallization of a monodisperse suspension of active repulsive spheres was not the goal of our study, and for this reason we have not  characterized the state of the system at such high densities. }\\

\textcolor{black}{ In Figure 2(a), 2(b) and 2(c) the P\'eclet is varied by changing  $k_B T$, $D_r$, and $F_a$, respectively. The three diagrams coincide when $Pe=16$ as the values of all the coefficients which define $\mathrm{Pe}$ are the same. A detailed discussion on a way to compare the different diagrams can be found in the supplementary material. Taking this into account, from the comparison between 1) the (a) and (b) state diagrams and 2) the (a) and (c) phase diagrams, we can isolate the influence of the following two physical properties:} 



\begin {enumerate}
\item \textcolor{black}{ Translational diffusion in the case where $D_r$ is reduced to increase $\mathrm{Pe}$, \ref{fig:WCADr}(b), is less than when $k_BT$ is reduced, \ref{fig:WCADr}(a), for a given $\mathrm{Pe}$ (leaving all else unchanged).}
\textcolor{black}{The phase diagrams for both WCA and PHS in figure \ref{fig:WCADr}(b) show a shift of the MIPS boundary towards higher $\rho$ and $Pe$ when compared to \ref{fig:WCADr}(a). This preference for the homogeneous state suggests that a greater effective translational diffusion hinders MIPS.}

\item 

\textcolor{black}{ When increasing $F_a$, the relative strength of the force leading to collisions compared to the repulsive force increases. This is not the case when $k_B T$ is reduced. Therefore for a given P\'eclet number, the particles are effectively softer when $F_a$ is increased. This is visible when comparing \ref{fig:WCADr}(c) to \ref{fig:WCADr}(a). When using the WCA potential, the softer potential, changing $F_a$ shifts the MIPS boundary to higher P\'eclet numbers and densities. However, there are fewer differences when using the harder PHS potential. This indicates that the WCA potential is not a good representation for hard disks in the range of parameters used in this work. The PHS potential seems to be strong enough to prevent any significant particle overlap in both methods. Note - Previous version very unclear.}


\end{enumerate}

 \textcolor{black}{
To study the structural features of the ABP suspensions, we compute the value of $\psi_6$ for each particle (eq.6, to get information on the local crystalline order) and on the entire system (averaging over all particles, to get information on the global crystalline order).}

\begin{widetext}

 \begin{figure}[h!]
  \centering
  \includegraphics[width=0.61\columnwidth]{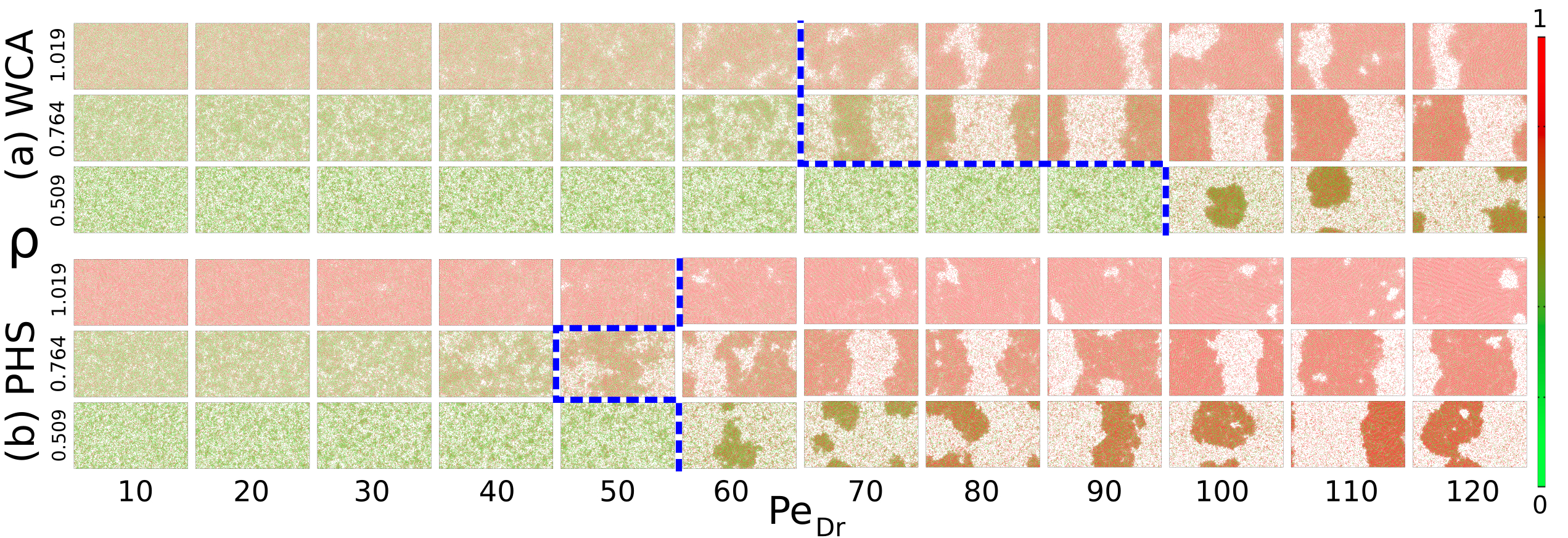} \includegraphics[width=0.38\columnwidth]{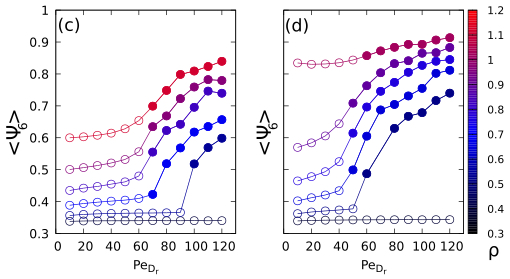}
  \caption{\textcolor{black}{(a) WCA and (b) PHS snapshots of the system in  steady-state at selected points of the $\rho$-$\mathrm{Pe}_{D_r}$ state diagram (as indicated in the vertical/bottom axes). The color-code  corresponds to the local $\psi_6 $order, ranging from 0 (low order, green) to 1 (high order, red) and the blue dashed line correspond to the boundary of MIPS in figure \ref{fig:WCADr} panel (b). Panels (c) WCA and (d) PHS represents global value of $\psi_6$ (averaged over all particles in the system) for all  points in the $\rho$-$\mathrm{Pe}_{D_r}$ state diagram. Empty and filled dots represent homogeneous and MIPS states, respectively, with the same criteria used in figure \ref{fig:WCADr} (see panel (b)).}
    \label{fig:PsiSnap}}
  \end{figure}
  
  \end{widetext}

   \textcolor{black}{
In figure \ref{fig:PsiSnap}(a) and (b) we show different snapshots for the PHS and WCA systems, respectively \textcolor{black}{(see supplementary material for a bigger version of this panels)}. The colour code reflects the value of $\psi_6$ for each particle, ranging from low (green) to high (red) local crystalline order. 
Independent of the interaction potential, we identify three main outcomes: 1) a homogeneous dilute phase of disordered particles (mostly  green particles); 2) a MIPS phase, characterised by a dense phase (red particles) and a dilute phase (green particles); 3) an homogeneous dense phase of mostly ordered particles (in red). 
Interestingly, for the WCA-ABP suspension, MIPS is shifted towards higher densities, due to a lower effective diameter arising from partial particle overlap\cite{rogel}.}

\textcolor{black}{Clearly, 
the stiffness of the interaction potential not only has a considerable effect on the shape of the phase diagram (as shown in Figure 2), but it also affects how particles are ordered inside the dense phase. In figure \ref{fig:PsiSnap}(c) (WCA) and (d) (PHS) we show how the global order parameter changes when varying $\rho$ and $\mathrm{Pe}_{Dr}$. At low densities, both systems are disordered, independent of the value of the activity $\mathrm{Pe}_{Dr}$. However, when increasing density, $\langle\psi_6\rangle$ starts increasing for values of $\mathrm{Pe}_{Dr}$ in close correspondence to the appearance of MIPS. }

\textcolor{black}{When comparing the two interaction potentials, WCA particles have a lower hexatic order than PHS ones at the same densities and $\mathrm{Pe}_{Dr}$ values. This is because WCA particles partially overlap, whereas hard disks must organize with a higher hexagonal order at high density.}

  \textcolor{black}{ 
It is interesting to note that $\psi_6$ is not a good indicator to locate the MIPS boundaries at least for high density systems, because its value strongly depends on the density, making it difficult to define a unique threshold value of $\psi_6$ that can help detect MIPS. Moreover, as we can see in figure \ref{fig:PsiSnap} panels c and d the difference in $\left \langle \psi_6 \right \rangle$ between MIPS and homogeneous states is not so clear at high densities. Some of the selected snapshots (corresponding to high densities and low $\mathrm{Pe}$ number, especially for the PHS potential) 
are characterised by a relatively large average value of $\psi_6$. However, these are regions of the state diagram where no MIPS has been detected by neither the local density nor the non-Gaussian parameter (see Figure \ref{fig:WCADr}). }

\subsection{Using the non-Gaussian parameter as a way to identify MIPS}

In Sec.A, we study the phase behaviour of soft-like (WCA) and hard-like (PHS) excluded-volume potentials, determining MIPS regions (via the local density). We explore how different ways of modifying the P{\' e}clet number can affect the shape and location of the MIPS boundary in the state diagram, depending on the stiffness of the interaction potential. MIPS, characterized by the appearance of high/low density regions, are identified via static properties such as the probability distribution function of the local density $P(\rho_0)$, and structurally  characterized by $\psi_6$. Although the static and spatial inhomogeneities are frequently used to identify MIPS, it is clear that MIPS states are also characterized by a very large dynamical heterogeneity: in high density regions, particles are almost stagnant; whereas in low density regions, some particles may move very fast during the 
short periods of time between collisions.

\textcolor{black}{The same behaviour is characteristic of metastable fluids, where dynamic heterogeneities are present. When studying colloidal suspensions, Weeks and coworkers \cite{weeks2000three} demonstrated that for purely diffusive particles, the distribution of particles displacements was Gaussian. However, deviations from a Gaussian start to appear when the system becomes metastable (e.g. over-compressed) and can be quantified by a non-Gaussian parameter $\alpha_2$ (which is exactly zero for a Gaussian distribution \cite{Rahman}). When approaching the glass transition packing fraction, the value of $\alpha_2$ increases. This is a signature of the fact that the system is characterized by spatially-correlated aggregates of fast particles in a ``sea'' of slow particles \cite{donati1,donati2,markus,kegel}}


Thus, when a fluid-like system is in a homogeneous state, the particle displacement is expected to be Gaussian\textcolor{black}{, even in the presence of activity}. Between collisions, the equation of motion has two terms, the diffusive term (which always yields Gaussian distributions) and the self-propulsion term. The random nature of particle collisions creates trajectories that are very similar to random 
flights after just a few collision events. Therefore, we expect that particle displacement should be Gaussian in homogeneous states, at least for times longer than the average times between collisions. However, when MIPS occurs, the system separates in phases of very different mobility which strongly affects the Gaussianity of the displacements distribution. This has inspired us to search for an alternative method to identify MIPS: 1) computing the probability distribution function of particle displacement, 2) we detect the  departures from Gaussianity, i.e. the two-dimensional non-Gaussian parameter $\alpha_2$, eq. (\ref{eq:nongauss}) (expanding an idea 
we proposed in  ref.\cite{rogel}). 
\begin{figure}[h!]
\centering
	\includegraphics[width=0.98\columnwidth]{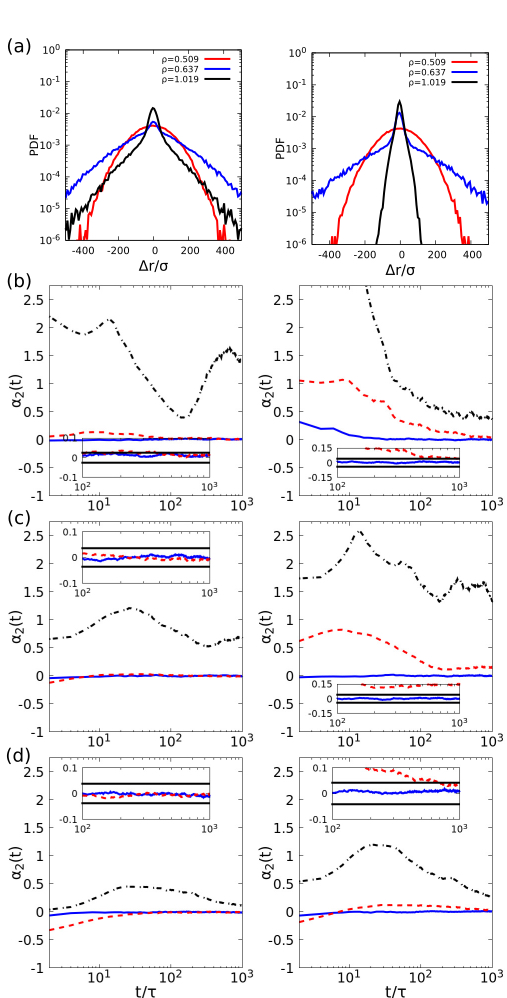}
  \caption{\textcolor{black}{(a) Steady state probability distribution functions of particle displacements at different densities and $\mathrm{Pe}_{D_r}$ for WCA (left) and PHS (right): $\rho=0.51$, $\mathrm{Pe}_{D_r}=20$ (red, homogeneous case), $\rho=0.64$, $\mathrm{Pe}_{D_r}=90$ (blue, MIPS at low packing fraction) and $\rho=1.02$, $\mathrm{Pe}_{D_r}=120$ (black, MIPS at high packing fraction), at the same lag time ($t=128$). }
  Time evolution of the 2D non-Gaussian parameter $\alpha_2$ for systems of ABP interacting with the WCA (left) and the PHS (right) potentials, at densities (b) $\rho=1.019$, (c) $\rho=0.764$ and (d) $\rho = 0.509$, and $\mathrm{Pe}_{D_r}$ values equal to $10$ (blue solid line), $60$ (dashed red line) and $120$ (dash-dot black line). All simulations start in a stationary condition. The inset panels show a zoom of the last steps of the simulations (from $t=100$ to $1000$). Black horizontal lines show the threshold used to establish MIPS ($10\sigma_\alpha$, see discussion after eq. (\ref{eq:nongauss})).
	 \label{fig:nonG}}
\end{figure}

\textcolor{black}{In Figure \ref{fig:nonG}(a), the probability distribution functions of particle displacements in the steady state are shown for ABPs interacting via the WCA (left-hand side) and PHS (right-hand side) potentials, at different values of $\mathrm{Pe}_{D_r}$ and density. When the system is in a homogeneous state ($\rho=0.509$, $\mathrm{Pe}_{D_r}=20$, in red), the PDF is clearly Gaussian, which is shown as a parabola in the semi-logarithmic plot. However, as soon as the system enters MIPS ($\rho=0.64$, $\mathrm{Pe}_{D_r}=90$, in blue, and $\rho=1.02$, $\mathrm{Pe}_{D_r}=120$, in black), the PDF has a non-Gaussian shape in which the contributions from the slow and fast-moving particles (those inside and outside the MIPS region, respectively), can be clearly detected.}

In figures \ref{fig:nonG} (b), (c) and (d), we show the time evolution of $\alpha_2$  for $\rho$ (from high (b) to low (d)) and $\mathrm{Pe}_{D_r}$ ranging from $10$ (blue continous line), $60$ (red-dashed line) to $120$ (black dashed-dotted line). 
The measurements are taken starting from initial configurations in the steady-state, 
for systems of particles interacting with the WCA (left-hand side) and PHS (right-hand side) potentials. In the insets, the thresholds of admissible non-Gaussianity are depicted as horizontal black lines, corresponding to ten times the size of fluctuations for a given system at the same density and in the absence of any activity. The choice of the factor $10$ is arbitrary, but appears to be a reasonable choice if we assume that the distribution of instantaneous values of $\alpha_2$ for a passive homogeneous system is Gaussian, and considering that, in a perfect Gaussian distribution, $99.7$\% of all $\alpha_2$ values should be included within the  $[-3\sigma_\alpha, 3\sigma_\alpha]$ interval. Excursions in the value of $\alpha_2$ of sizes larger than $10\sigma_\alpha$ 
are thus indicative of non-Gaussianity in the particle displacements and, therefore, of dynamical heterogeneity and MIPS. \textcolor{black}{Note that, although the systems are in the steady state, the instantaneous value of $\alpha_2$ is not constant. Other equilibrium systems, such as gels of associating polymers \cite{ramirez2018molecular} show a similar trend, in which alpha becomes greater than zero for a given time, before returning to zero at very long times.}

For $\mathrm{Pe}_{D_r}=10$ (blue solid lines in Fig. \ref{fig:nonG}) none of the three states for both WCA and PHS ends in a MIPS state (see Fig. \ref{fig:WCADr}), and the evolution of $\alpha_2$ is within the threshold limits either at all times or after a short ``equilibration'' time, equal to a few characteristic times between collisions. For $\mathrm{Pe}_{D_r}=60$ (red dashed lines in Fig. \ref{fig:nonG}) none of the three states for the WCA potential is in a MIPS state (see Fig. \ref{fig:WCADr}) and $\alpha_2$ is also within the threshold of non-Gaussianity. However, for the PHS potential, all three states present MIPS and $\alpha_2$ shows clear signs of non-Gaussianity. At density $\rho = 0.509$, the excursions of $\alpha_2$ outside the limits are short-lived and reach moderate values, which is consistent with the fact that the point lies at the boundary of the MIPS region in the phase diagram of PHS (see Fig. \ref{fig:WCADr}). Finally, for $\mathrm{Pe}_{D_r}=120$ (black dash-dot lines in Fig. \ref{fig:nonG}), all three states for both WCA and PHS show MIPS (see Fig. \ref{fig:WCADr}), and the evolution of $\alpha_2$ is clearly non-Gaussian from very early times (as shown  in Fig. 5 of the the Supplementary Information)

{\color{black}{Therefore, the results obtained with $\alpha_2$ to establish whether the system is in a homogeneous or MIPS state coincide with those obtained when computing the local density.}}
The calculation of the non-Gaussian parameter is a straightforward method to identify MIPS states, and computationally far less demanding than other methods such as computing probability distributions of local densities using Voronoi cells. 
In terms of computational requirements, both methods need the system to reach  steady-state. On the one hand,  
in order to compute  $P(\rho_0)$, one needs good statistics (long runs and uncorrelated configurations). 
On the other hand, $\alpha_2$ is very easy to calculate (even included in LAMMPS \cite{LAMMPS}), and MIPS can be detected without the need of very long runs and for large numbers of particles. In MIPS states, $\alpha_2$ rapidly becomes very large.
In addition, we anticipate that the advantages of using the non-Gaussian parameter to detect MIPS over other static methods will be even clearer when simulating systems of ABP in three dimensions.

\subsection{Dynamical features of MIPS}

{\color{black}{
In the previous section, we show that when the system enters a MIPS phase, the non-Gaussian parameter is clearly non-zero, as expected for a system separating into a dilute region of fast particles and a dense region of slow particles. The appearance of dynamical heterogeneities is a characteristic feature of supercooled liquids approaching the glass transition\cite{berthier-biroli}. Similarly, several studies have been performed on trying to understand the non-equilibrium glass transitions of active particles \cite{henkes1,angelini2,dijkstra,wysocki6,fily7,flenner,bi12,mandal13,ding14}}}

{\color{black}{
In ref.\cite{Ludovic}, the authors studied the dynamics of a WCA-ABP binary mixture. They computed the effective diffusion coefficient as a function of packing fraction and  found that, at low effective temperatures, the diffusion coefficient increased with increasing persistence time (or activity).
To determine the glass transition line, they fitted the diffusion coefficient to a Vogel–Fulcher-like dependence on the packing fraction, $\ln D = A + B (\phi - \phi_c)$, where $A$, $B$ and $\phi_c$ (the glass transition packing fraction) were fitting parameters. From the data, the authors inferred that, at low effective temperature, the glass transition packing fraction monotonically increased with activity. }}

\begin{figure}[h!]
\centering
\includegraphics[width=1\columnwidth]{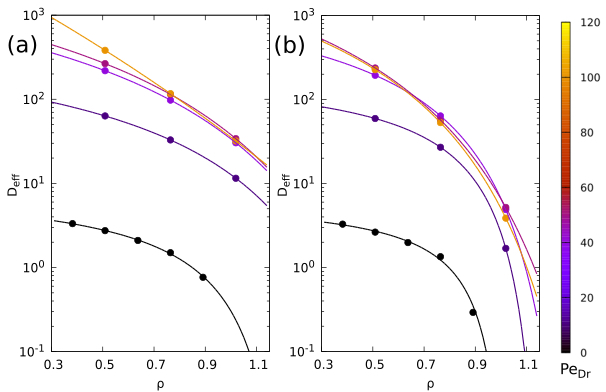}
\caption{Diffusion coefficient vs $\rho$ for (a) WCA and for (b) PHS for $\mathrm{Pe}_{D_r}=0$ (black), $10$ (black-violed), $40$ (light-violed), $80$ (orange). Lines are the Vogel-Fulcher model fit for this data. Note that  this fit is performed with the results of simulations shown in figure \ref{fig:PsiSnap}. \label{fig:Deff}}
\end{figure}

{\color{black}{As in ref.\cite{Ludovic}, }}
we study the dynamics by computing the mean-square displacement and extracting the effective diffusion coefficient from its long-time behaviour. 
In figure \ref{fig:Deff} we represent the diffusion coefficient {\color{black}{ as a function of density}}
for passive particles (black lines), particles in homogeneous states (purple lines)
and particles in MIPS states (orange lines), {\color{black}{when the P\'eclet number is varied by changing $D_r$}}.

When dealing with passive particles {\color{black}{(black lines)}}, unsurprisingly, the values of their diffusion coefficients are lower than those of active particles {\color{black}{(coloured lines)}}. The diffusion coefficients {\color{black}{of passive particles}}, when represented as a function of density, {\color{black}{can be fitted by}} a Vogel-Fulcher expression\cite{Ludovic}. Interestingly, the stiffer potential {\color{black}{(right-hand side in Fig.5)}} reduces the effective diffusion at higher densities when aggregation becomes important. For the WCA potential {\color{black}{(left-hand side in Fig.5)}}, the diffusion coefficient is significantly higher than for PHS at high densities, this stems from the softer nature of the interaction which allows a certain degree of particle overlap and an easier particle motion.
 
When dealing with active particles, effective diffusion is always higher than for passive counterparts. Moreover, the effective diffusion for very low densities is identical in both potentials. At higher densities, a stiffer potential like PHS is more efficient at preventing particle overlap and thus has a lower effective diffusion coefficient than a softer potential like WCA. The Vogel-Fulcher equation seems to be a valid model to describe the decay of the effective diffusion coefficient with density {\color{black}{(see table in the Supplementary Information)}}, even for active systems\cite{Ludovic}, except in the highest density/P\'eclet number regime. In this parameter range, the Vogel-Fulcher fit yields the equivalent of a glass transition density $\rho_0$ which is too large to have any physical meaning {\color{black}{(see table in the Supplementary Information)}}, even considering the larger degree of overlap between WCA particles.

{\color{black}{Therefore, we conclude that the PHS-ABP diffusion coefficient can be fitted by a Vogel-Fulcher expression in a more meaningful way than the WCA-ABP one (see table 2 in the supplementary information). Since this relationship is used to describe the behaviour of fragile glasses we suggest that, as in Ref.\cite{weitz}, a suspension of PHS-ABP behaves like a fragile glass, whereas this behaviour is less clear when particles are softer (WCA-like).}}

It is important to note that the mean-square displacement in the presence of MIPS  {\color{black}{(orange lines in Figure 5)}} is the result of the average between fast and slow phases. 
{\color{black}{When dealing with passive systems close to the glass transition the fraction of fast particles is extremely low. 
While the fast(dilute)/slow(dense) phases seem to be more equally present in WCA-ABP, in PHS-ABP the MIPS state is characterized by small regions of low (fast) density immersed in a sea of high (slow) density. Therefore, the contribution of fast particles in the mean-square displacement is lower for PHS. For this reason, the VFT fits, which are supposed to work for passive colloidal glasses, are less meaningful for active colloids at high densities and activities, independent of the stiffness of the interaction potential.}}



\section{Conclusions}

We have studied a two dimensional suspension of repulsive ABPs, interacting via two different repulsive potentials: WCA and a stiffer PHS potential. To characterise their structural features, we have studied their phase behaviour when varying the P{\'e}clet number in several ways. 

We have found that, in addition to $\mathrm{Pe}$ and $\rho$, two other parameters have an effect on the onset of MIPS: the relative strength of the effective translational diffusion with respect to the self-propulsion force; and the relative strength of the  interaction forces with respect to the self-propulsion.

The translational diffusion hinders the emergence of MIPS, preventing its formation at low $\rho$ and also shifting the MIPS boundary to higher $\mathrm{Pe}$.

When the self-propulsion force is strong enough compared to the repulsive excluded volume interactions, the softness of the potential plays a role in the state diagram. When interacting via WCA, the overlapping of particles in the dense region induces a shift of the MIPS phase boundary to higher $\rho$. This effect can be avoided by choosing the stiffer PHS potential, at the cost of having to use a smaller time-step.
Therefore, different repulsive potentials might lead to differences in the phase behaviour and this is particularly the case when simulating ``hard'' active matter particles. 

For increasing $\rho$, we identify the following states, independent of the interaction potential: dense phases of slow particles with low density clusters, bands and bubbles of fast particles,  
in this order. For stiffer potentials like PHS, these phenomena can be observed at lower densities. \textcolor{black}{Note that the observed percolating bands always appear along the shortest dimension of the box, showing that this particular morphology could be the consequence a finite size effect.}

To better characterise these morphologies, we compute the crystalline order, both local and global. 
\textcolor{black}{The stiffness of the interaction potential not only has a considerable effect on the state diagram, but it also affects particle order inside the dense phase. When comparing the two interaction potentials, WCA-ABP have a lower hexatic order than PHS ones at the same densities-$\mathrm{Pe}_{Dr}$ values. The reason is that WCA particles are allowed to partially overlap, differently from PHS, that at high density organize in an hexagonal. To conclude, $\psi_6$ cannot be used as a good indicator to locate the MIPS boundaries.}

We suggest an alternative way to better identify the MIPS boundaries, based on particles' dynamics instead of local density. 
Since particle displacements should be Gaussian in homogeneous states, we inspect the value of the two-dimensional non-Gaussian parameter $\alpha_2$ to detect MIPS. Our results show that both static and dynamic methods yield the same MIPS boundary. Using the non-Gaussian parameter is easier, less CPU intensive and faster than other static methods, such as computing the local density. 

\textcolor{black}{Finally, we have studied the system's dynamics and found that the effective diffusion coefficient can be fitted by a Vogel-Fulcher equation, as for over-compressed colloidal suspensions. However, this fit is less meaningful for active colloids at high densities and activities, independent of the stiffness of the interaction potential.}



\section*{Conflicts of interest}
There are no conflicts to declare. 

\section*{Acknowledgements}

The authors acknowledge funding from Grant
FIS2016-78847-P and ID2019-105343GB-I00  of the
MINECO and the UCM/Santander PR26/16-10B-2.
Francisco Alarc\'on acknowledges support from the ``Juan de la Cierva''
program (FJCI-2017-33580).
Raul Martinez acknowledge funding from MICINN (Ministerio de Ciencia e Innovaci\'on, Spain) FPI grant BES-2017-081108.
 The authors acknowledge the
computer resources and technical assistance provided by the
Centro de Supercomputaci\'on y Visualizaci\'on de Madrid
(CeSViMa) and from the Red Espa\~nola de Supercomputacion (RES)  FI-2020-1-0015 and 
FI-2020-2-0032.

\balance


\providecommand*{\mcitethebibliography}{\thebibliography}
\csname @ifundefined\endcsname{endmcitethebibliography}
{\let\endmcitethebibliography\endthebibliography}{}


\section{Supplementary information: Characterization of MIPS in a suspension of repulsive Active Brownian Particles through dynamical features}

{\color{black}{
\subsection{Comparison between different ways of changing $\mathrm{Pe}$}

In this section, we attempt to clarify the differences between the different ways of varying $\mathrm{Pe}$. We start by showing equations (1) and (2) of the main text:
\begin{align}
\label{eq:motion}
& \dot{\vec{r}}_i = \frac{D_t}{k_B T} \left( - \sum_{j\neq i} \nabla V(r_{ij}) +  F_a \, \vec{n}_i \right) + \sqrt{2D_t} \, \vec{\xi}_i, \\
& \dot{\theta}_i = \sqrt{2D_r}\, \xi_{i,\theta}.
\end{align}
The $\mathrm{Pe}$ number is defined in the text as:
\begin{equation}
\mathrm{Pe} = \frac{3 v \, \tau_r}{\sigma}= \frac{3 F_a D_t}{\sigma k_B T D_r},
\label{eq:Pe}
\end{equation}
which measures the ratio between the active force and the reorientation of particles. Looking at eq. \ref{eq:Pe}, the $\mathrm{Pe}$ number can be modified by changing $k_B T$, $F_a$ or $D_r$. Although the route of changing $F_a$ is more easily achieved experimentally, all three different options have been used in simulations in the past. 
In the main text, we compare the ensuing phase diagrams when $\mathrm{Pe}$ is modified by changing $k_BT$ and $D_r$ (Figs 2(a) and 2(b) in the main text, respectively). We assert that the difference between both diagrams lies in the effective relative strength of the translational diffusion term ($\sqrt{2D_t} \, \vec{\xi}_i$ in equation \ref{eq:motion}). This is counter-intuitive since the value of $D_t$ is kept constant in both sets of simulations. Here, we attempt to clarify this assertion by means of an example.

We start from a particular value of $\mathrm{Pe}=\mathrm{Pe}_0$ where all the coefficients used in the simulation are exactly the same in both diagrams (in our work, this corresponds to $\mathrm{Pe}_0=16$), and consider a value $\mathrm{Pe}=\lambda \text{Pe}_0$. In the case where $\mathrm{Pe}$ is modified by changing $k_B T$, the equation in discrete form reads:
\begin{align}
\label{ekt}
& \Delta\vec{r}_i = \lambda\frac{D_t}{k_B T} \left( - \sum_{j\neq i} \nabla V(r_{ij}) +  F_a \, \vec{n}_i \right)\Delta t + \sqrt{2D_t \Delta t} \, \vec{\xi}_i,\nonumber \\
& \Delta\theta_i = \sqrt{2D_r \Delta t}\, \xi_{i,\theta},
\end{align}
whereas in the case that $\mathrm{Pe}$ is modified by changing $D_r$, it reads
\begin{align}
\label{edr}
& \Delta\vec{r}_i = \frac{D_t}{k_B T} \left( - \sum_{j\neq i} \nabla V(r_{ij}) +  F_a \, \vec{n}_i \right)\Delta t + \sqrt{2D_t \Delta t} \, \vec{\xi}_i, \nonumber\\
& \Delta\theta_i = \sqrt{2D_r \Delta t / \lambda}\, \xi_{i,\theta},
\end{align}

If we rescale the time step in Eq. \ref{edr} by setting $\Delta t'\rightarrow \Delta t/\lambda$, it yields:
\begin{align}
\label{drrees}
& \Delta\vec{r}_i = \lambda\frac{D_t}{k_B T} \left( - \sum_{j\neq i} \nabla V(r_{ij}) +  F_a \, \vec{n}_i \right)\Delta t' + \sqrt{\lambda · 2D_t \Delta t'} \, \vec{\xi}_i,\nonumber \\
& \Delta\theta_i = \sqrt{2D_r \Delta t'}\, \xi_{i,\theta},
\end{align}

As we are studying the behavior of the system in the steady state, any change of the time scale is not relevant in determining the eventual formation of MIPS. So, we can see that the difference between equation \ref{ekt} (obtained by changing $k_B T$) and equation \ref{drrees} (obtained by changing $D_r$) is reflected in the effective transnational diffusion, that is $\lambda D_t$ in the second case.

Since, in our diagrams, we have chosen a reference value of $\mathrm{Pe}_0=16$, which is below the threshold to obtain MIPS, $\lambda>1$ in the MIPS region and we can say that the translational diffusion term is relatively greater near the MIPS boundary when $\mathrm{Pe}$ is modified by changing $D_r$ than in the case of changing $k_B T$, this being the only relevant difference between both diagrams.

An analogous reasoning leads us to assert that the most relevant effect when comparing the diagrams obtained when $\mathrm{Pe}$ is modified by changing $k_B T$ and changing $F_a$ is that the excluded-volume repulsion force is effectively weaker in the second case.}}

{\color{black}{
\subsection{WCA and PHS state diagrams}

In Figure 2-main text, we show three state diagrams for each potential (WCA and PHS), each phase diagram obtained: (a) changing $k_BT$ (while keeping $\epsilon=1)$, (b) changing $D_r$ but leaving $D_t$ unchanged (thus not using Stokes-Einstein), and (c) changing $F_a$.

In each panel, the system undergoes motility induced phase separation (MIPS, filled symbols in Figure 2-main text) when increasing both density and P\'eclet number.
Whereas, the system is in an homogeneous phase at low density and low P\'eclet number (empty symbols in Figure 2-main text).

In the main text, we already suggested that the different parameters chosen to vary the P\'eclet mostly affect the transition to MIPS, rather than its bulk.
To support this statement, we now represent all 6 phase diagrams in the same figure.
\begin{figure}[h!]
\centering
    \includegraphics[width=0.7 \columnwidth]{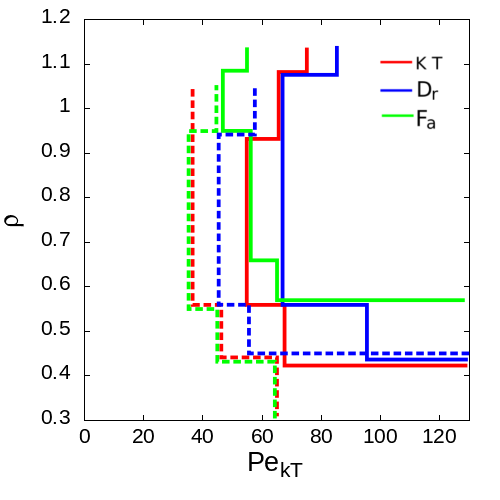} 
\caption{ Phase diagrams' boundary for PHS (dashed lines) and WCA (continuous lines) for  \label{fig:allPD}}
\end{figure}

Figure\ref{fig:allPD} shows the MIPS boundaries as obtained by means of the non-Gaussian parameter, for the ABP-WCA (continuous lines) and the ABP-PHS (dashed lines), when varying $k_BT$ (in red), $D_r$ (in blue) and $F_a$ (in green).
Even though the figure is a bit crowded, it is clear that while the exact definition of the MIPS boundaries depends on the method chosen to vary Peclet, the bulk region of the MIPS phase is not affected.}}

\subsection{Structural Features}

\begin{figure}[h!]
\centering
    \includegraphics[width=0.8 \columnwidth]{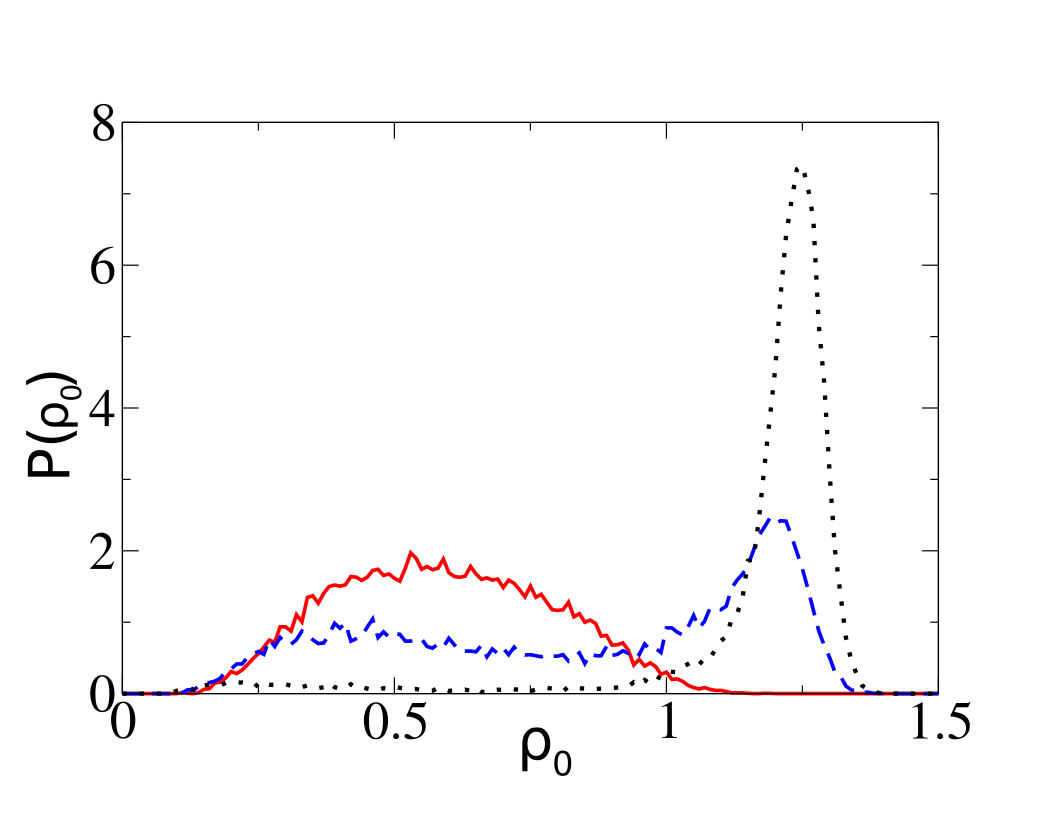} 
\caption{Probability Distribution Function of the local density for a system of spherical Active Brownian particles interacting with a WCA potential. 
The continuous red line corresponds to the homogeneous case, $\rho=0.51$($\phi=40$), $\mathrm{Pe}=20$; the dashed blue line corresponds to a low density and high activity system, close to the MIPS, $\rho=0.64$ ($\phi=50$), $\mathrm{Pe}=90$; and the dotted black line corresponds to a high density and activity system, $\rho=1.02$ ($\phi=80$), $\mathrm{Pe}=120$. \label{fig:Pvoro}}
\end{figure}

One of the goals of the main text is to study state diagram for both potentials and see the differences depending of the softness. As we can see in others references like Ref.\cite{Stenhammar}, one possible criteria to identify MIPS with is via local density distribution function, as explained in the methods section of main text. In Fig. \ref{fig:Pvoro}, we represent the system in a homogeneous (the red continuous line), and MIPS phase (the blue-striped and black-dotted lines). The criteria for this approach follow: the distribution is uni-modal for homogeneous behavior, bi-modal for MIPS behavior and it is unclear if there is two maxima when close to the MIPS boundary of the state diagram. This method is based in a static property of the system as we have discussed in main text, and we propose a new way using dynamical parameters of the system.\\

\subsection{Dynamical Features}

In the main text we introduce a new way to identify MIPS based on the Non-Gaussian parameter (Eq. 7 in main text). This parameter shows the deviation of the probability distribution function (PDF) of particle displacements in two dimensions from a Gaussian. Some examples of this distribution are showed in Fig. 4 of the main text. There we compute the steady state PDF of the three selected systems at the same lag time  ($t/\tau = 128$, system units) averaged over many starting configurations.

Note that if a PDF is a Gaussian distribution in a log-linear axis, such as the one used in figure 4 in main text (panel a), the distribution has an apparent parabolic shape.

By analysing the non-Gaussianity of the probability distribution function (PDF) of particle displacements, we establish the presence of slow and fast regions in the system, that can are related to the formation of the MIPS state.

The homogeneous state (red line) clearly follows a Gaussian distribution, whereas the boundary state (blue) and the MIPS state deviate from Gaussianity, both for the WCA and the PHS potential. At the boundary of MIPS, the PDF seems to split into two Gaussian-like distributions, one for fast moving particles (those in the low density regions) and another one, highly peaked at zero, for slow moving particles (those trapped in the MIPS region).




{\color{black}{
\subsection{Meaning of $\sigma_\alpha$}

In the absence of activity, the probability distribution function (PDF) of particle displacements along axes $x$ or $y$ is always Gaussian (check Fig 4a in the main text) and the system is always in a homogeneous state. The non-Gaussian parameter $\alpha_2$ has been proposed as an easy-to-measure observable to check if the PDF is Gaussian \cite{kumar2006nature}. If the displacement distribution remains Gaussian at all times, $\alpha_2$ should be always very close to zero. However, it is not exactly zero due to the fluctuations related to the finite size of the system and the fluctuations in density. In order to characterize the size of those fluctuations, we run simulations of each system at a given density and $\mathrm{Pe}=0$, and collect the instantaneous values of $\alpha_2$ at equilibrium. Then, we calculate the standard deviation of $\alpha_2$, which we define as $\sigma_\alpha$. If the distribution of instantaneous values of $\alpha_2$ is Gaussian, then it is known that the interval $[-3\sigma_\alpha, 3\sigma_\alpha]$ includes 99.7\% of all possible excursions of $\alpha_2$ away from zero. In our method to detect MIPS, we hypothesize that if a system with $\mathrm{Pe}>0$ remains homogeneous, then $\alpha_2$ will fluctuate around zero and the size of the fluctuations will be very similar to the corresponding passive case. However, when the system enters MIPS, $\alpha_2$ becomes greater than zero (note that it becomes greater than zero only temporarily; at very long times, when all particles have explored all possible states, both inside MIPS and outside MIPS, $\alpha_2$ will go back to zero). When MIPS is clear (at high densities and high $\mathrm{Pe}$), $\alpha_2$ becomes clearly greater than zero and it is easy to detect. However, when the system is close to the MIPS boundary in the state diagram, $\alpha_2$ only increases slightly, and we need to establish some criteria to detect MIPS. Here, we choose the following criterion: any excursion of $\alpha_2$ that goes beyond $10\sigma_\alpha$ will be a sign of non-homogenous displacements and the appearance of MIPS states. The factor $10$ is fully arbitrary, and it is chosen because the probability that a homogeneous system at the same density shows a value of $\alpha_2$ greater than $10\sigma_\alpha$ tends to zero. Our results show that this choice of $\sigma_\alpha$ is a reasonable one when the goal is to detect MIPS.}}

{\color{black}{
\subsection{Vogel-Fulcher-Tamman fitting parameters}
 
 In what follows, we present the fitting parameters obtained when fitting the effective diffusion coefficient as a function of density, as reported in Figure 5 of the main text for both repulsive potentials: WCA (left-hand side) and PHS (right-hand side)}
 
 Note that the values of $\rho$’ (which is supposed to be related to the glass transition density) take non-physical values for the WCA potential as soon as MIPS appears ($\mathrm{Pe}>60$). Such extremely large densities (for example, above 5) have no meaning since there is no way a system of WCA particles can be reasonably packed at those densities.}
 
\begin{table}[h!]
    \centering
    \begin{tabular}
    {|>{\color{blue}}c| >{\color{blue}}c|>{\color{blue}}c|>{\color{blue}}c|}
    $\mathrm{Pe}$ & $A_{WCA}$ & $B_{WCA}$ & $\rho_{WCA}$' \\ 
    \hline
0   & 2.21612       & 0.890521      & 1.25013 \\
10	& 7.00822		& 3.90604		& 1.87473 \\ 
20	& 8.22868		& 5.13315		& 1.98743 \\ 
30	& 8.82889		& 5.59404		& 2.01109 \\ 
40	& 9.72677		& 7.08431		& 2.14124 \\ 
50	& 10.2519		& 7.79958		& 2.17904 \\ 
60	& 10.1073		& 6.59472		& 2.02504 \\ 
70	& 31.6134		& 150.755		& 6.37311 \\ 
80	& 25.3437		& 77.2729		& 4.51103 \\ 
90	& 36.5183		& 188.766		& 6.71897 \\ 
100	& 32.8438		& 162.439		& 6.54804 \\ 
110	& 34.6641		& 191.96		& 7.18215 \\ 
120	& 35.8994		& 204.484		& 7.33301 \\ 

    \end{tabular}
    \caption{Fitting parameters of the VFT relation.}
    \label{tab:table_WCA}
\end{table}

\begin{table}[h!]
    \centering
    \begin{tabular}{|>{\color{blue}}c| >{\color{blue}}c|>{\color{blue}}c|>{\color{blue}}c|}
    $\mathrm{Pe}$ & $A$ & $B$ & $\rho$' \\ 
    \hline
0   & 1.73164   & 0.375296  & 1.01796 \\
10	& 5.49945	& 1.00892	& 1.22178 \\ 
20	& 6.46029	& 1.35961	& 1.26864 \\ 
30	& 7.23691	& 1.80904	& 1.32708 \\ 
40	& 8.08591	& 2.53682	& 1.40942 \\ 
50	& 11.0476	& 7.00222	& 1.7639 \\ 
60	& 10.9699	& 7.14034	& 1.78665 \\ 
70	& 13.1184	& 12.0143	& 2.06341 \\ 
80	& 9.50974	& 4.44673	& 1.56797 \\ 
90	& 8.68816	& 3.52368	& 1.50861 \\ 
100	& 10.4906	& 5.80438	& 1.65429 \\ 
110	& 7.15466	& 1.67549	& 1.30497 \\ 
120	& 9.98273	& 4.95099	& 1.57988 \\ 
    \end{tabular}
    \caption{PHS}
    \label{tab:table_PHS}
\end{table}



\newpage
\begin{figure*}[hbtp]
\centering
    \includegraphics[width=0.9 \columnwidth]{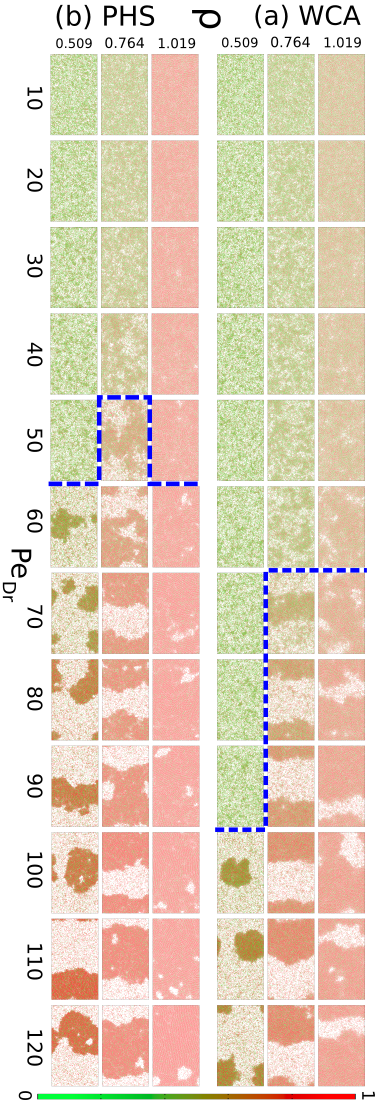} 
\caption{(a) WCA and (b) PHS snapshots of the system in  steady-state at selected points of the $\rho$-$\mathrm{Pe}_{D_r}$ state diagram (as indicated in the vertical/bottom axes). The color-code  corresponds to the local $\psi_6 $order, ranging from 0 (low order, green) to 1 (high order, red) and the blue dashed line correspond to the boundary of MIPS in figure \ref{fig:WCADr} panel (b).}
\end{figure*}

\end{document}